\documentclass[a4paper,10pt]{swaluw}
\usepackage{graphicx}
\newcommand{\halfskip}{\vskip 0.5\baselineskip \noindent}
\newcommand{\be}{\halfskip \begin{equation}}
\newcommand{\ee}{\end{equation} \halfskip}
\newcommand{\nskip}{\vskip \baselineskip \noindent}
\begin{document}

\authorrunning{E. van der Swaluw et al.}
\title{Interaction of High-Velocity Pulsars with Supernova Remnant Shells}
\author{E. van der Swaluw \inst{1,2}, A. Achterberg \inst{1}, Y.A. Gallant
\inst{1,3}, T.P. Downes \inst{4} and R. Keppens \inst{5}}
\institute{Astronomical Institute, Utrecht University, P.O.Box 80000, 3508 TA 
Utrecht, The Netherlands
\and
Dublin Institute for Advanced Studies, 5 Merrion Square, Dublin 2, Ireland
\and
Service d'Astrophysique, CEA Saclay, 91191 Gif-sur-Yvette Cedex, France
\and
School of Mathematical Sciences, Dublin City University, Glasnevin, Dublin 9,
Ireland
\and
FOM-Institute for Plasma Physics Rijnhuizen, P.O. Box 1207, 3430 BE Nieuwegein, The Netherlands 
}
\date{}
\abstract{
Hydrodynamical simulations are presented of a pulsar wind emitted by a 
supersonically moving pulsar.
The pulsar moves through the interstellar medium or, in the more interesting case, 
through the supernova remnant created at its birth event. 
In both cases there exists a three-fold structure consisting of the wind termination shock,
contact discontinuity and a bow shock bounding the pulsar wind nebula. 
Using hydrodynamical simulations we study the behaviour of the pulsar wind nebula
inside a supernova remnant, and in particular
the interaction with the outer shell of swept up
interstellar matter and the blast wave surrounding the remnant. 
This interaction occurs when
the pulsar breaks out of the supernova remnant. We assume the remnant is in the Sedov stage
of its evolution. Just before break-through, the Mach number
associated with the pulsar motion equals ${\cal M}_{\rm psr} = 7/\sqrt{5}$, {\em independent}
of the supernova explosion energy and pulsar velocity. 
The bow shock structure is shown to survive this break-through event.
}
\titlerunning{Interaction of High-Velocity PSRs with SNR shells}
\maketitle

\section{Introduction}

\noindent A supernova explosion of a massive star will result in an expanding 
supernova remnant (SNR).  In some cases the fossil of the progenitor star
is a pulsar moving at high velocity. Even though the precise physical mechanism
responsible for imparting a large kick velocity to single radio pulsars at birth
has not been identified, observations of the pulsar distribution with respect to
the mid-plane of the galaxy indicate that they are born with a velocity in the range
$V_{\rm psr} \sim 100-1000$ km/s (Harrison et al. 1992; Lyne \& Lorimer 1994).
A similar range of values is obtained from a sample of SNR-pulsar associations
(Frail et al. 1994).

The expansion of a SNR is decelerated due to mass-loading by swept up
interstellar medium (ISM) or by material from a progenitor wind.
As the pulsar moves with a constant velocity it will ultimately
break through the SNR shell. Two observed systems are often presented as
an illustration of this scenario:
\nskip
{\it CTB80}: in this supernova remnant the pulsar
{\it PSR B1951+32} is located (in projection) 
just inside the outer edge of the remnant. The spectral index of
the synchrotron emission in the vicinity of the pulsar system 
indicates that there is a plerionic nebula around the pulsar, see for
example Strom (1987) and Strom \& Stappers (2000).
\nskip
{\it G5.4-1.2}: 
in this case 
the pulsar is located well outside the supernova remnant.
At radio frequencies an emission bridge appears to connect  the pulsar {\it B1757-24} 
and the associated pulsar wind nebula (PWN) with the supernova remnant 
(Frail \& Kulkarni 1991), suggesting a physical association between the
supernova remnant and the pulsar. It should be pointed out, however, 
that a new upper limit on the proper motion of {\it B1757-24} (Gaensler \& Frail 2000)
puts the component of the pulsar velocity in the plane of the sky at
$V_{\perp \rm psr} \le 600$ km/s for an assumed distance of 5 kpc. This leads
to a discrepancy between the characteristic pulsar age, obtained from its
spin period derivative ($P/2 \dot{P} \sim 16$ kyr), and the dynamical age 
obtained from the offset distance $R_{\rm psr}$ from the center of {\it G5.4-1.2}
($R_{\rm psr}/V_{\rm psr} \geq 39$ kyr).  
\nskip

Both systems are clearly brightened at radio wavelengths near the position of the
pulsar, and it has been suggested that the associated pulsar wind is rejuvenating 
the radio emission from the SNR shell by the injection of fresh
relativistic electrons (Shull et al. 1989). In this paper we will investigate the 
hydrodynamical aspects of the interaction between a pulsar wind and a SNR shell.

Most pulsars have a lifetime ($10^{6} - 10^{7}$ yr) which is much larger than the age
$< 10^{4}$ yr of a SNR in the Sedov phase. 
Therefore pulsars will remain visible long after the associated SNR has dissolved
into the interstellar medium.
The pulsar then moves as an isolated pulsar through the interstellar medium, and
can form a pulsar wind nebula bow shock system.
A typical example of such a system is the {\em Guitar Nebula} around {\it PSR B2224+65}
which has been detected both in H$\alpha$ (Cordes et al. 1993) and in 
X-rays (Romani et al. 1997), but which
has no associated SNR. 

In this paper we consider the case where the pulsar's kick velocity is sufficiently
high so that it leaves the supernova remnant while it is still in the Sedov stage.  
We describe three different stages in the evolution of the pulsar-SNR system: 
(1) the stage where the PWN/bow shock resides inside the SNR, (2) 
the PWN/bow shock breaking through the
shell of the SNR and (3) the stage where the PWN/bow shock moves through the 
ISM.

\section{Physics of a PWN bow shock inside a SNR}

\subsection{Dynamics of the pulsar/SNR system}

In rapidly rotating (young or recycled) pulsars, it is believed that
a pulsar wind carries away most of the spindown luminosity, 
\halfskip
\[
	L = I \Omega \dot{\Omega} \; , 
\]
\halfskip
of a pulsar with rotation period $P = 2 \pi/ \Omega$ and moment of inertia $I$. 
This relativistic wind is presumably generated in the pulsar magnetosphere,
and accelerates electrons, positrons and possibly nuclei to ultra-relativistic
speeds. 

The pulsar wind blows a bubble or pulsar wind nebula (PWN) into the surrounding medium. 
This PWN is
initially located well within the interior of the SNR created at the birth of the neutron
star. During the free expansion stage of the SNR evolution the typical expansion speed
of the stellar ejecta as determined by the mechanical energy $E_{\rm snr}$ released
in the explosion and the ejecta mass $M_{\rm ej}$,
\halfskip
\[
	V_{\rm ej} \sim 
	\sqrt{\frac{10}{3} \: \frac{E_{\rm snr}}{M_{\rm ej}}} \sim
	10\; 000 \; {\rm km/s} \; ,
\]
\halfskip
is generally much larger than the kick velocity of the pulsar. As a result the
PWN is located relatively close to the center of the SNR at this stage. In this
regime, the results of our earlier, spherically symmetric simulations (van der
Swaluw et al. 2001) are expected to remain approximately valid.
Only when the SNR expansion slows down as it enters the Sedov stage after some
$\sim 500-1\; 000\;\;{\rm yr}$, a situation is possible where the pulsar position becomes 
strongly excentric with respect to the SNR.

The Sedov stage of SNR expansion lasts until internal (radiative) cooling 
becomes important. The SNR then enters the so-called pressure-driven snowplow (PDS) 
stage.
The relevant transition time is calculated by Blondin et al. (1998):
\be
       t_{\rm PDS}= 2.9\times 10^4 \: E_{51}^{4/17} n_0^{-9/17} \; {\rm yr.}
\ee
Here $E_{51}$ is the explosion energy $E_{\rm snr}$ in units of $10^{51}$ ergs
and $n_0$ denotes the hydrogen number density in the ISM.     

We will describe the physics of a pulsar wind interaction with the shell 
of a SNR in the Sedov stage. 
The results presented 
below only apply for certain range of values for the pulsar velocity 
$V_{\rm psr}$. Equating the distance traveled by the pulsar,
\halfskip
\[
       R_{\rm psr} = V_{\rm psr} t \; ,
\]
\halfskip
with the radius for a SNR embedded in a homogeneous interstellar medium
of density $\rho_{\rm ism}$ in the Sedov stage,
\be
\label{Sedov}
       R_{\rm snr} \simeq 1.15 \: 
       \left(\frac{E_{\rm snr}}{\rho_{\rm ism}} \right)^{1/5}
       \: t^{2/5} \; ,
\ee
\halfskip        
one gets the crossing time for the pulsar:
\be
\label{crosstime}
       t_{\rm cr} = 1.27 \: 
       \left({E_{\rm snr}\over{\rho_{\rm ism} V_{\rm psr}^5}}\right)^{1/3} \; ,
\ee
which is
\be
      t_{\rm cr} \: \simeq \: 
      1.4 \times 10^4 \; E_{51}^{1/3} V_{\rm 1000}^{-5/3}n_0^{-1/3}
       \;
       {\rm yr.}
\ee       
Here $V_{1000}$ is the velocity of the pulsar in units of 1,000 km/sec, and
$n_{0}$ the hydrogen number density of the ISM in units of ${\rm cm}^{-3}$.  
The requirement $t_{\rm cr} \le t_{\rm PDS}$ yields the minimum velocity 
a pulsar needs in order to break out of the SNR while the latter is still in the Sedov
stage:
\be       
       V_{\rm psr} \ge 650\; n_0^{2/17} E_{51}^{1/17}
       \; {\rm km/s.}
\ee
Although this is a rather high value, it is still in the range of values 
observed by Harrison et al. (1992).

\begin{figure}
\resizebox{\hsize}{!}
{\includegraphics{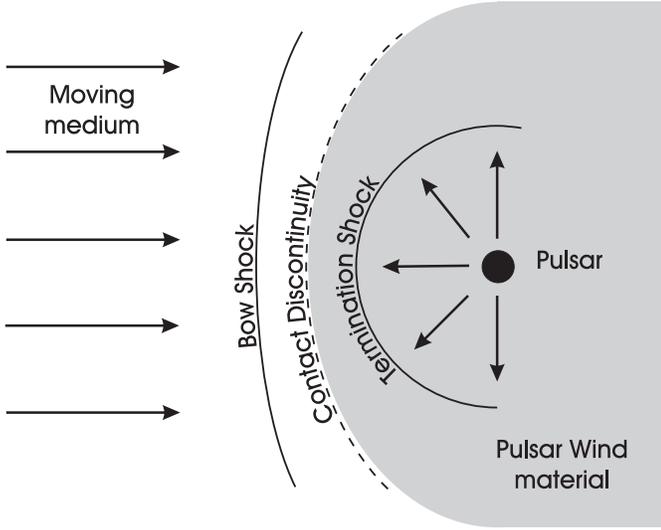}}
\caption{Configuration of the pulsar wind nebula moving through a uniform 
medium as seen in the rest frame of the pulsar.}
\end{figure}

One can use the Rankine-Hugoniot relations to determine the pressure just 
behind the Sedov-Taylor blast wave bounding the
SNR (assuming a strong shock and a gas with specific heat ratio $\gamma = 5/3$)
\be
\label{strongRH}
	P_{\rm sh} = \mbox{$\frac{3}{4}$} \: \rho_{\rm ism} \: 
	V_{\rm snr}^{2} \; ,
\ee	
where
\halfskip
\[
	V_{\rm snr} \equiv \frac{{\rm d} R_{\rm snr}}{{\rm d} t} 
	= \mbox{$\frac{2}{5}$} \: \frac{R_{\rm snr}}{t}
\]
\halfskip
is the SNR expansion speed.
Using this expression together with expression (\ref{crosstime}) for the crossing 
time and the Sedov solution (\ref{Sedov}), one can derive the following 
equations valid  at the moment of break-through. 
The speed of the pulsar is related to the SNR expansion speed by
\be
	V_{\rm psr} = \mbox{$\frac{5}{2}$} \: V_{\rm snr} \; ,
\ee
while the material in the shell behind the SNR blast wave moves with a velocity
\be
	V_{\rm sh} = \mbox{$\frac{3}{4}$} \: V_{\rm snr} \; .	 
\ee
This corresponds to a relative speed between pulsar and post-shock material
equal to
\be
	V_{\rm rel} \equiv V_{\rm psr} - V_{\rm sh} = \mbox{$\frac{7}{4}$} \:
	V_{\rm snr} \; .
\ee
The density in the shell is roughly the post-shock density for a strong shock, 
$\rho_{\rm sh} = 4 \rho_{\rm ism}$, so the
Mach number ${\cal M}_{\rm psr}$ of the pulsar motion through the shell 
material satisfies
\be
	{\cal M}_{\rm psr} = \frac{V_{\rm rel}}
	{\sqrt{\gamma P_{\rm sh}/4 \rho_{\rm ism}}}
	= \frac{7}{\sqrt{5}} \approx 3.13 \; .
\ee	
Since the proper motion of the pulsar is supersonic with respect to the
surrounding medium the outer rim of the PWN will become a-spherical and a bow
shock will form.

\subsection{Pulsar Wind}

A pulsar wind is believed to consist of an ultra-relativistic, cold
flow with a large bulk Lorentz factor, e.g. $\Gamma_{\rm w} \ge 10^{3}$
in the case of the Crab (Gallant et al. 2002). The cold wind is terminated 
by a termination shock which thermalizes the flow, leading to a 
relativistically hot downstream state with sound speed $s \sim c/\sqrt{3}$.
Following Kennel \& Coroniti (1984), the luminosity of the pulsar wind  at 
the wind termination shock, $R_{\rm ts}$ is given by a combination of particle 
and magnetic luminosity:
\be
\label{fluxes}
	L = 4\pi \Gamma_{\rm w}^{2} n_{\rm w} R_{\rm ts}^{2}  
	m c^{3}(1+\sigma) \; ,
\ee   
where $n_{\rm w}$ is the proper density in the wind, $m$ the mean mass
per particle and $\sigma$ is the ratio of Poynting flux to plasma kinetic 
energy flux. In this paper we consider an unmagnetised pulsar wind, i.e. 
$\sigma = 0$. The typical pressure behind the ultra-relativistic termination 
shock is (e.g. Kennel \& Coroniti 1984)
\be
\label{pwpress}
	P_{\rm ts} \approx \mbox{$\frac{2}{3}$} \: 
	\Gamma_{\rm w}^{2} n_{\rm w} m c^{2} =
	\frac{L}{6 \pi R_{\rm ts}^{2} c} \; .
\ee
The first equality is approximate because of deviations of sphericity of the 
pulsar wind region, induced by the proper motion of the pulsar.
The shocked pulsar wind material is separated from material that has gone
through the bow shock by a contact discontinuity.
Because of the high internal sound speed, both in the pulsar wind material 
and in the material that has passed through the bow shock, and because of
the small size of the region between the
termination shock and bow shock, this region 
can be considered to be isobaric to lowest approximation.
Using the Rankine-Hugoniot jump conditions at the stagnation point at the head
of the bow shock surrounding the PWN, the pressure equals:
\be
\label{inside}
	P_{\rm bs} = (5{\cal M}_{\rm psr}^2 - 1) P_{\rm sh} / 4 = 
	9 \: \rho_{\rm ism} V_{\rm snr}^{2} \; .
\ee
After the pulsar has broken through the shell the pulsar wind is completely
confined by the ram pressure of the cold ISM and the stagnation-point pressure,
using the Rankine-Hugoniot relations again, drops to
\be 
\label{outside}
       P_{\rm bs} = \mbox{$\frac{3}{4}$} \rho_{\rm ism} V_{\rm psr}^2 = 
       \mbox{$\frac{75}{16}$} \:  \rho_{\rm ism} V_{\rm snr}^2 \; ,
\ee      
a pressure reduction by roughly 50\% as the pulsar leaves the SNR. 
The fact that the region between termination shock and bow shock is almost
isobaric implies
\halfskip
\[
	P_{\rm bs} \approx P_{\rm ts} \; .
\] 
\halfskip
This condition, together with Eqn. (\ref{pwpress}), determines the stand-off distance of the 
pulsar wind termination shock as
\be
\label{termrad2}      
       R_{\rm ts} = \eta \: \left(\frac{L}{6 \pi \rho_{\rm ism} V_{\rm psr}^2
       c}\right)^{1/2} \; ,
\ee
where the numerical factor $\eta$ is determined by Eqns (\ref{inside}) or (\ref{outside}). 
This parameter takes the value $\eta = 5/6 \approx
0.83$ when the pulsar is still just inside the SNR, and $\eta =
2/\sqrt{3}\approx 1.15$ when the
pulsar moves through the ISM.  The radius $R_{\rm ts}$ is also the typical 
stand-off distance of the bow shock, which is always close to the termination 
shock at the head of the pulsar wind nebula.

These expressions allow us to calculate the relative size of the pulsar wind
to the supernova remnant at the moment of break-through.
From the expression (\ref{crosstime}) for the crossing time one has
\halfskip
\[
	R_{\rm snr}(t_{\rm cr}) = V_{\rm psr} t_{\rm cr} =
	13.6 \; E_{51}^{1/3} V_{\rm 1000}^{-2/3}n_0^{-1/3}
	\; {\rm pc} \; .
\]
\halfskip
The termination shock radius is of order
\halfskip
\[
	R_{\rm ts} \simeq 2.8 \times 10^{-3} \: \eta \: 
	L_{36}^{1/2} n_{0}^{-1/2} V_{1000}^{-1} \; {\rm pc} \; ,
\]
\halfskip
here $L_{36} = L/(10^{36} \; {\rm erg/s})$.
Note that the size of the SNR shell is much larger then the size of
the PWN. For this reason we will neglect the curvature of the
SNR blast wave and perform a hydrodynamical simulation
where the pulsar moves with a Mach number of ${\cal M}_{\rm psr}= 7/\sqrt{5} \simeq 3.13$ 
through the post-shock flow of a strong plane-parallel shock,
ultimately crossing this shock into the unshocked medium.

\section {Numerical simulations of the PWN bow shock}
                    
\subsection{Simulation Method}
                  
\begin{figure}
\resizebox{\hsize}{!}
{\includegraphics[scale=0.7,angle=270]{NewTable.epsi}}
\end{figure}

We simulate a pulsar wind using the Versatile Advection Code 
\footnote{See http://www.phys.uu.nl/\~{}toth/}
(VAC), a  general-purpose software package developed initially by G.  
T\'oth at the Astronomical Institute in Utrecht (T\'oth  1996; T\'oth \& Odstr\u cil  1996).
The configuration is depicted in figure 1, showing both
shocks which are of interest: the pulsar wind termination shock and the 
bow shock bounding the PWN. This system is assumed to be axially symmetric 
around the direction of motion of the pulsar. Out of the several choices for 
discretizing the equations of hydrodynamics in conservative form available in 
VAC,  we use a shock-capturing, Total-Variation-Diminishing Lax-Friedrich 
scheme (T\'oth \& Odstr\u cil, 1996). Our simulations have been performed
in two dimensions, using a cylindrical coordinate system.  

\begin{figure}
\resizebox{\hsize}{!}{\includegraphics{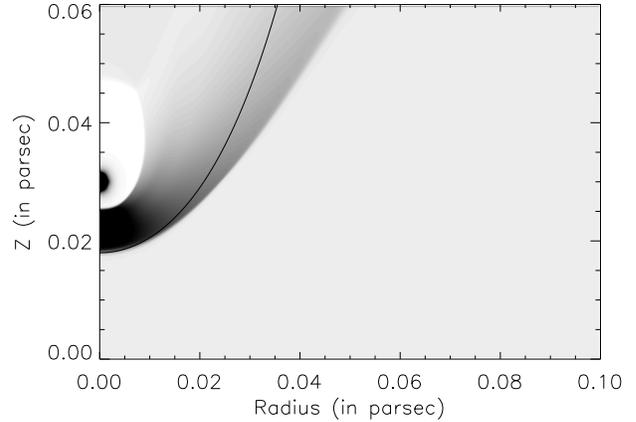}}
\caption{ Comparison between the numerical result for the bow shock with a low
Mach Number (${\cal M} = 7/\sqrt{5} \approx 3.13$) with the profile given by 
Wilkin (1996) for ${\cal M} \gg 1$ (solid line). The grey scale represents 
the pressure.}
\end{figure}

\subsection{Initialising a pulsar wind} 
 
The current version of the hydrodynamics code uses one fluid with a single
equation of state in the computational domain. For simplicity, and accuracy in
the non-relativistic part of the flow, 
we use an ideal fluid with specific heat ratio $\gamma = 5/3$.  
The pulsar wind is modeled by continuously depositing thermal energy 
at a constant rate $L$ (the spindown luminosity) in a small volume, 
together with an associated mass injection $\dot M_{\rm pw}$.
The hydrodynamics code itself then develops a thermally driven
wind with terminal velocity $v_\infty$  which is subsequently randomised at a
termination shock. 
In order to increase the computationally efficiency, we take the mechanical 
luminosity $L$ and mass deposition rate $\dot M_{\rm pw}$ such that the 
terminal velocity of the wind as determined from these two parameters is much 
larger than the other velocities of interest:
\halfskip
\[
       v_\infty =\sqrt{2 L/\dot M_{\rm pw}} .
\]
\halfskip       
This choice will result in the correct global behaviour of the PWN. 
This method is similar to the method as
described in van der Swaluw et al. (2001). We employ a non-uniform grid
with highest resolution near the pulsar in order to fully resolve the pulsar 
wind region. The non-uniform grid is chosen in such a way that the difference in
size between adjacent grid cells is always less then 5 \% .
The radius of the termination shock,
given by Eqn. (\ref{termrad2}) in the relativistic case, is replaced by its 
non-relativistic equivalent,
\halfskip
\[
	R_{\rm ts} \approx \eta \: \left( \frac{L}
	{2 \pi \rho_{\rm ism} V_{\rm psr}^{2} v_{\infty}} \right)^{1/2} \; ,
\]
\halfskip 
and will have roughly the correct value when $v_{\infty} \approx c$. 

\subsection{Steady PWN bow shock in a uniform medium}

Our calculations are performed in the pulsar rest frame.
The pulsar wind nebula is allowed to evolve in a uniform medium, moving at a
constant speed $V_{\rm psr}$ at large distances from the pulsar.
This medium represents the interior of the supernova remnant (shocked ISM)
close to the blast wave.
The velocity $V_{\rm psr}$ is supersonic with respect to the internal 
sound speed of the medium so that a bow shock develops around the PWN. 
We let the hydrodynamics code evolve the system until the large-scale 
flow is steady. 

\begin{figure}
\resizebox{\hsize}{!}{\includegraphics{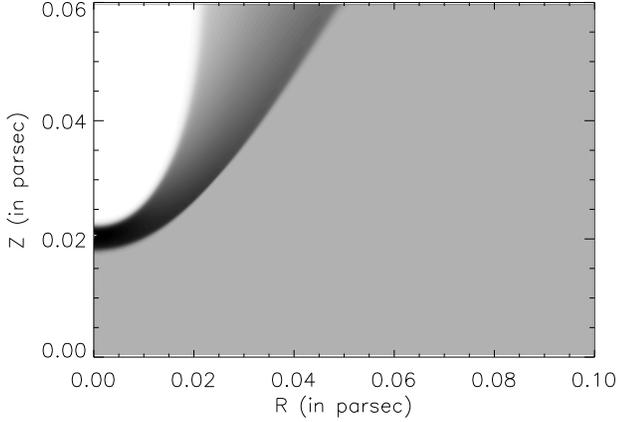}}
\caption{ Gray-scale representation of the density distribution of a PWN and
bow shock with the parameters as denoted in table 1.}
\end{figure}

\begin{figure}
\resizebox{\hsize}{!}{
\includegraphics{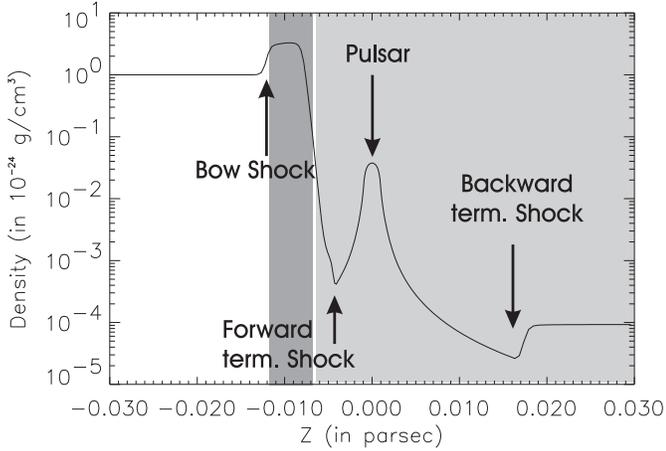}}
\caption{ Density profile of the PWN and bow shock along the z-axis of 
Figure 3. The location of the Pulsar, forward and backward 
termination shocks and of the bow shock are indicated. The pulsar wind region 
is shaded light gray, the region containing shocked interstellar material is 
shaded a darker gray.}
\end{figure}

\begin{figure}
\resizebox{\hsize}{!}{\includegraphics{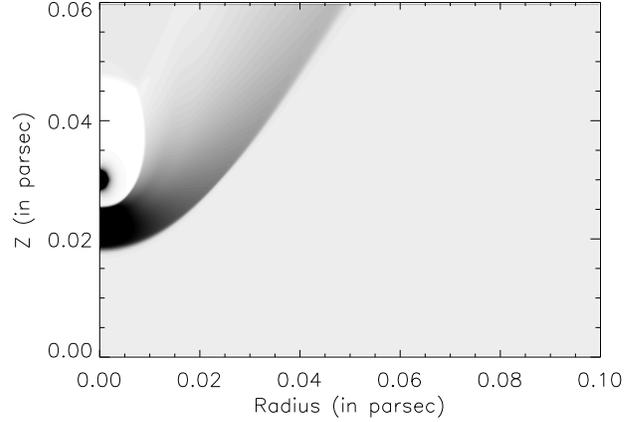}}
\caption{Gray-scale representation of the pressure distribution of a PWN bow 
shock with the same parameters as in Figure 3.}
\end{figure}

\begin{figure}
\resizebox{\hsize}{!}
{\includegraphics{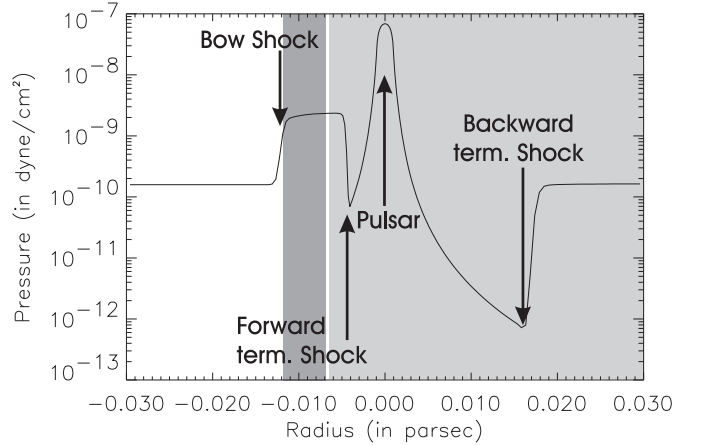}}
\caption{Pressure profile of a PWN and bow shock along the z-axis of 
figure 5. Shading is as in Figure 4.}
\end{figure}

This simulation has been performed with parameters as denoted in table 1.
In order to determine when the system is steady, we employ the prescription 
of T\'oth et al.(1998). This prescription compares all $N_{\rm var}$ flow 
variables at time $t_{n}$, denoted by $U^{n}_{\alpha i}$ at grid point $i$,
with their values at the previous time $t_{n-1}$. 
We then calculate the residual ${\tt Res}$ defined as
\be       
       {\tt Res} \; \equiv \; 
       \sqrt{\displaystyle {1\over N_{\rm var}}\sum_{\alpha=1}^{N_{\rm var}}
       {\sum_{i} \left[U^{n+1}_{\alpha i} - U^{n}_{\alpha i} \right]^2\over 
       {\sum_{i} \left[ U^{n}_{\alpha i} \right]^{2}}}} \; ,
\ee
and halt the calculation when ${\tt Res}$ has a value less than a 
predetermined critical value, typically ${\tt Res} = 10^{-4}$.

Wilkin (1996) has given an analytical equation for the geometry of a stellar 
wind bow shock. His solution, for the distance $r$ to the wind source in terms
of the polar angle $\theta$ with respect to the symmetry axis, reads:
\be
	\frac{r(\theta)}{R_{0}} \; \equiv \;
	{1\over\sin\theta} \: 
	\sqrt{\displaystyle 3 \: \left(1-{\theta\over\tan\theta} \right)} \; .
\ee
Here $R_{0} \approx R_{\rm bs}$ is the stand-off distance of the bow shock
on the symmetry axis ($\theta = 0$).
We compare our morphology with Wilkin's result, where we equate 
$R_0$ to the stand-off distance of the bow shock in the simulations. 
This is depicted in figure 2. The asymptotic cone of Wilkin's solution 
is seen to be significantly narrower than that of the bow shock in the
simulation. This is because his solution balances only the ram pressures 
of the wind and the ambient medium, i.e. it corresponds to the limit 
${\cal M}_{\rm psr} \gg 1$, while in our case the Mach number is moderate: 
${\cal M}_{\rm psr} \approx 3.13$.

The figures 3 and 4 show density profiles of the PWN bow shock of a pulsar
moving through a uniform medium. One can see the contact discontinuity, midway between 
the termination shock and the bow shock, separating  the shocked pulsar wind material and the much
denser shocked ISM. The synchrotron emission of the plerionic
PWN is expected to come from the shocked pulsar wind material, whereas the
material swept up by the bow shock can show up as H$\alpha$ emission.
 
Figure 5 shows the pressure distribution and figure 6 shows the pressure 
profile along the symmetry axis.
One can see the pulsar wind, originating at the pulsar position $z = 0$,
and its termination shock at $z \approx -0.005$ pc ahead of the pulsar
in the direction of motion, and at $z \approx 0.017$ pc behind the pulsar. 
The bow shock bounding the PWN is located at $z \approx -0.012$ pc.
The region between the pulsar wind termination shock and the bow shock
is almost isobaric. As shown by van der Swaluw et al. (2001), this is 
also the case for a PWN around a stationary pulsar located at
the center of the SNR. 

\begin{figure}
\resizebox{\hsize}{!}{\includegraphics{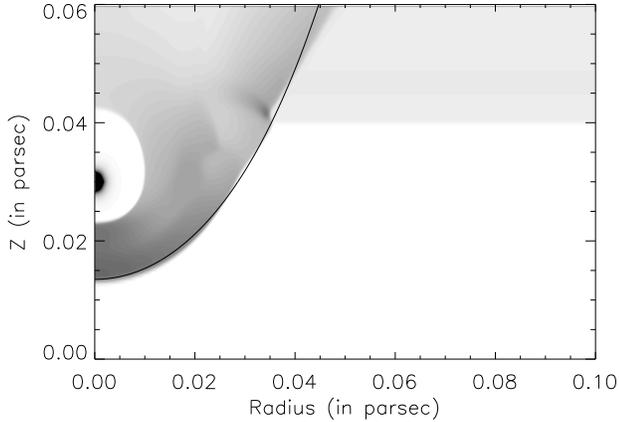}}
\caption{ Pressure distribution of a PWN bow shock with parameters as 
denoted in table 1. Here the PWN is interacting 
with the shell of the SNR, modeled as a plane-parallel shock. The profile
given by Wilkin (1996) for ${\cal M}\gg 1$ (solid line) agrees well with the
bow shock profile in the ISM.}
\end{figure}

\subsection{Interaction of the PWN with a shock}

In this section we present results of the break-through of 
the PWN bow shock when it crosses the shell of its supernova remnant. 
The simulation is performed once more in the rest frame of the pulsar. 
We initialise a steady-state configuration of the PWN bow shock as described 
above, with the same parameters as denoted in table 1. Next we use the 
Rankine-Hugoniot relations to initiate a strong shock front 
(with Mach number ${\cal M}\sim 100$) near the bottom (low $z$) end of the 
grid, with which the pulsar wind bow shock then catches up in accordance with 
equations (7)-(10). As stated in section 2, the PWN bow shock is much smaller 
than the radius of the SNR, so we can safely approximate the SNR blast wave as 
a plane-parallel strong shock.

At the end of the simulation, when the strong shock is almost at the upper 
boundary of the grid, numerical instabilities arise. Therefore we stop the 
simulation after the configuration as shown in the figures 7-9, when the 
influence of the numerical instabilities are not influencing the solution 
too strongly.

Figures 7 and 8 show the system, some time after the SNR shock has passed the 
head of the bow shock.  Figure 7 shows the pressure distribution and figure 8 
the density distribution.
Figure 9, which compares the density profile along the $z$-axis before and after the
passage of the supernova blast wave, shows that the pulsar wind nebula expands, 
roughly by a factor 1.5, after it leaves the SNR. This reflects the reduction 
in the confining (ram-)pressure, as calculated in Section 2. 
Figure 7 also shows that the bow shock shape in the ISM now closely fits the 
analytical result of Wilkin (1996). Once the pulsar wind bow shock has crossed 
the supernova blast wave, its Mach number with respect to the interstellar 
medium is much larger than unity. This agreement is all the more remarkable 
given that strictly speaking, the assumptions under which Wilkin's solution is 
derived (geometrical thinness and complete mixing of the post-shock fluids) 
are not fulfilled in pulsar bow-shock nebulae (Bucciantini \& Bandiera 2001).

\begin{figure}
\resizebox{\hsize}{!}{\includegraphics{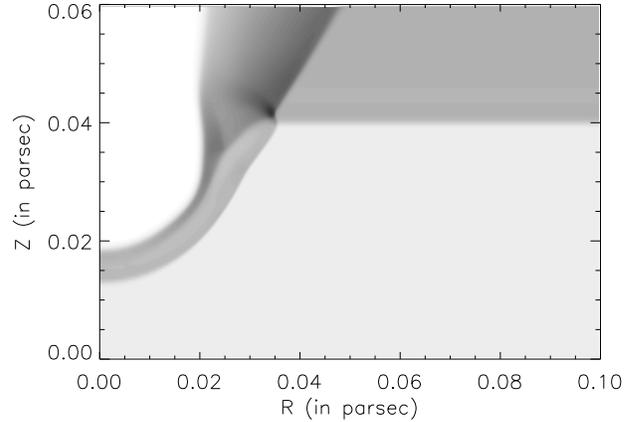}}
\caption{ Density distribution of the PWN bow shock-SNR interaction corrsponding
to the pressure distribution of Figure 7}
\end{figure}

\begin{figure}
\resizebox{\hsize}{!}
{\includegraphics{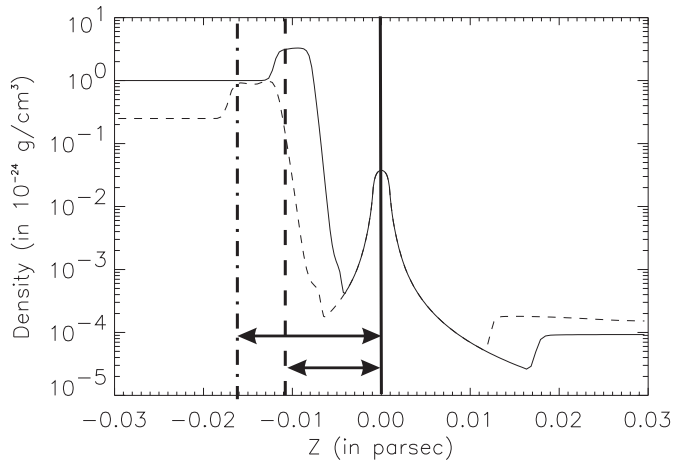}}
\caption{ Density profile of the PWN bow shock along the z-axis.
In this figure the pulsar's position corresponds to $Z = 0$. The solid
curve denotes the density profile before interacting with the SNR shock. The
dashed curve denotes the density profile well after the SNR shock has passed the head
of the bow shock. The dashed line and the dash-dot line give the position
of the bow shock around the PWN before and after the passage of the
supernova blast wave. The PWN expands roughly by a factor 1.5.}
\end{figure}

During the interaction between the pulsar wind and the shell of the SNR,
the PWN bow shock and the SNR blast wave intersect.
This intersection produces an 
additional pressure gradient which results in an accumulation of 
mass. The resulting pressure and density enhancements can be 
seen as dark spots at the region of intersection in the figures 7 and 8. 
When the bow shock moves through the shell of the remnant it encounters the
unshocked ISM. The ambient density is reduced by a factor 4, which results
in a similar density reduction behind the bow shock.

\section{Conclusions}

We have considered the case of a pulsar wind breaking through the shell of 
a SNR in the Sedov-Taylor stage. Only high-velocity pulsars
reach the edge of the SNR while the latter is still in the Sedov stage of its 
evolution.
At  the moment of break-through, the ratio of the pulsar velocity and SNR 
expansion speed is $V_{\rm psr}/V_{\rm snr}=5/2$, and the Mach number 
associated with the pulsar motion equals ${\cal M}_{\rm psr}=7/\sqrt{5}$. 
These conclusions are  {\it independent} of the explosion energy $E_{\rm snr}$ 
or the pulsar speed $V_{\rm psr}$.

Our simulations show that the break-through of the PWN
does not lead to a significant disruption. The reduction of stagnation 
pressure by about 50\% leads to a moderate expansion of the PWN where its radius
increases by a factor $\sim 1.5$.
The latter result was also obtained analytically in Eqns. (13)-(15).

There is good agreement between our numerical results and analytical estimates,
based on pressure balance arguments, for the size of the bow shock surrounding
the PWN.  The only clear indication of the interaction between the PWN bow shock 
and the SNR (Sedov-Taylor) blast wave is a density and pressure enhancement at the
intersection of these two shocks.

In a subsequent paper we will consider the effects of the energetic particles which
were injected by the pulsar wind into the surroundings; some preliminary 
results of this investigation can be found in van der Swaluw et al. (2002).

\begin{acknowledgements}
The Versatile Advection Code was developed as part of the {\em Massief Parallel Rekenen}
(Massive Parallel Computing) program funded by the Dutch Organisation for Scientific Research (NWO). 
The authors thank Dr. S. Falle (Dept. of Applied Mathematics, University
of Leeds) for his assistance. EvdS is currently supported by the European 
Commission under the TMR programme, contract number ERB-FMRX-CT98-0168.
Y.A.G. acknowledges
support from the Netherlands Organisation for Scientific Research
(NWO) through GBE/MPR grand 614--21--008, and a Marie Curie
Fellowship from the European Commission, contract number HPMFCT-2000-00671.  
  
\end{acknowledgements}

\end{document}